\documentclass[showpacs,aps,prl,twocolumn,superscriptaddress,10pt]{revtex4-2}
\usepackage{graphicx} 
\usepackage{bm}
\usepackage{color}

\usepackage{amsmath}
\usepackage{amssymb}
\usepackage{enumerate}
\usepackage{xspace}
\usepackage{mathrsfs}
\usepackage{mathptmx}
\usepackage[utf8]{inputenc}

\newcommand{\sS}{\ensuremath{\mathbf{S}}\xspace}

\newcommand{\smmu}{\ensuremath{\mu_{\mathrm{s}}}\xspace}

\newcommand{\smB}{\ensuremath{\mathbf{B}}\xspace}

\newcommand{\sms}{\ensuremath{\mathbf{S}}\xspace}

\newcommand{\Jij}{\ensuremath{J_{ij}}\xspace}
\newcommand{\Jnn}{\ensuremath{J_{ij}^{\mathrm{nn}}}\xspace}
\newcommand{\Jnnn}{\ensuremath{J_{ij}^{\mathrm{nnn}}}\xspace}

\newcommand{\IrMn}{\ensuremath{\text{IrMn}_3}\xspace}
\newcommand{\Neel}{N\'eel\xspace}

\newcommand{\etal}{\textit{et al}\xspace}

\begin{document}

\title{The origin of exchange bias in multigranular non-collinear IrMn$_3$/CoFe thin films}

\author{Sarah Jenkins}
\email{sarah.jenkins@york.ac.uk}
\affiliation{Department of Physics, University of York, York, YO10 5DD, UK}
\author{Roy.~W.~Chantrell}
\affiliation{Department of Physics, University of York, York, YO10 5DD, UK}
\author{Richard.~F.~L. Evans}
\email{richard.evans@york.ac.uk}
\affiliation{Department of Physics, University of York, York, YO10 5DD, UK}

\begin{abstract}
Antiferromagnetic spintronic devices have the potential to greatly outperform conventional ferromagnetic devices due to their ultrafast dynamics and high data density. A challenge in designing these devices is the control and detection of the orientation of the anti-ferromagnet. One of the most promising ways to achieve this is through the exchange bias effect. This is of particular importance in large scale multigranular devices. Previously, due to the large system sizes, only micromagnetic simulations of exchange have been possible, with an assumed a distribution of antiferromagnetic anisotropy directions and grain size. Here, we use an atomistic model where the distribution of antiferromagnetic anisotropy directions occurs naturally and where the exchange bias occurs due to the intrinsic disorder in the antiferromagnet. We perform large scale simulations of exchange bias, generating realistic values of exchange bias. We find a strong temperature dependence of the exchange bias in agreement with experimental observations, approaching zero at the blocking temperature of the antiferromagnet. We find that the experimentally observed increase in the coercivity at the blocking temperature occurs due to the superparamagnetic flipping of the antiferromagnet during the hysteresis loop cycle. We find a large discrepancy between the exchange bias predicted from a geometric model of the antiferromagnetic interface indicating the importance of grain edge effects in multigranular exchange biased systems. The grain size dependence of the shows the expected peak due to a competition between the superparamagnetic nature of small grains and reduction in the statistical imbalance in the number of interfacial spins for larger grain sizes. Our simulations confirm the existence of single antiferromagnetic domains within each grain. The model gives insights into the physical origin of exchange bias and provides a route to developing optimised nanoscale antiferromagnetic spintronic devices.
\end{abstract}

\maketitle

\section{Introduction}
The development of novel anti-ferromagnetic spintronic devices could create information storage with a high data density, ultrafast dynamics and a robustness to external magnetic fields not seen in conventional ferromagnetic devices~\cite{Jungwirth2016AntiferromagneticSpintronics}. In these anti-ferromagnetic spintronic devices, the anti-ferromagnet is used to store and transmit information. The most significant problem in the development of these devices is the control and detection of the orientation of the antiferromagnet as they are impervious to applied magnetic fields. Electrical stimulation and detection of the orientation of an antiferromagnet has been measured~\cite{Godinho2018ElectricallyAntiferromagnet,Zelezny2018SpinDevices,Asa2020ElectricalEffect}, although the read-out signals are still small at room temperature. One of the most promising ways of controlling and detecting the magnetisation of antiferromagnetic materials is through the exchange bias effect. The exchange bias effect occurs when a ferromagnet (FM) is coupled to an anti-ferromagnet (AFM) and causes a shift of the magnetic hysteresis loop of the FM. The exchange bias effect has been used to obtain 180 degree switching using spin orbit torques but the mechanism for the switching is still not understood. To obtain full control of the AFM we need to fully understand the exchange bias effect. This is of particular importance in large scale granular AFM media as used in devices. 

Exchange Bias occurs due to uncompensated spins in the AFM at the FM/AFM interface, where these spins cause a unidirectional field on the FM. The exchange bias is determined from the number of uncompensated spins as \cite{Meiklejohn1957NewAnisotropy}:
\begin{equation}
    |B_{\mathrm{EB}}| = \frac{n_{\mathrm{ir}}J_{int}}{\mu_{\mathrm{FM}}n_{\mathrm{FM}}}
    \label{eq:EBnun}
\end{equation}

where $n_{FM}$ is the number of ferromagnetic atoms and $\mu_{FM}$ is the magnetic moment of the FM atoms. Since the discovery of exchange bias, many models have been developed to try and understand the origin of these uncompensated interface spins~\cite{Nowak1999MagnetizationRotation,Schulthess1998ConsequencesFilms,Bea2008MechanismsFilms,Stiles1999ModelBilayers,Mauri1987SimpleSubstrate}. Most of these models were based on the idea that the uncompensated spins occurred due to AFM domains or impurities. Recently, a new model has been proposed by Jenkins \etal~\cite{JenkinsEB2020}. They proposed a natural model of exchange bias for $\gamma$-\IrMn / CoFe bilayers which included a realistic $3Q$ tetrahedral spin structure in the antiferromagnet. The model gave accurate values for the exchange bias loop shift and the increase in coercivity for a single grain. They found the origin of the exchange bias originates from the natural structural disorder in IrMn, creating a small statistical imbalance in the number of interfacial spins. Their model is the first to explain the origin of exchange bias without the need for AFM domains or impurities. So far, the model has only been used for a single grain structure (8nm $\times$ 8nm $\times$ 8nm) whereas in realistic devices the IrMn is comprised of multiple grains and the systems are tens of times larger.

In multigranular thin films the exchange bias can be predicted from the grain size distribution. O'Grady \etal~\cite{OGrady2010AFilms} assumed the anisotropy of the AFM ($K_{AF}$) to be constant and therefore said the energy barrier within a grain is dictated by its volume ($V$) ~\cite{OGrady2010AFilms}. The probability of a grain switching is therefore dependent on the volume as: 

\begin{equation}
\tau^{-1} = f_0 \exp \left(-\frac{K_{AF}V}{k_BT}\right) ,
\label{eq:switch}
\end{equation}

\noindent where $\tau$ is the relaxation time, $k_B$ is the Boltzmann constant and $T$ is the temperature.

However, in reality not all the grains will set. If the grains are larger than the set volume ($V_{\mathrm{set}}(T)$) the relaxation time will be too long to set the uncompensated interface moment of these grains, and they will not be aligned with the FM layer. Furthermore, if the volume is too small the grains will be superparamagnetic at room temperature and therefore also not contribute to the exchange bias. Therefore only grains with grain volume $V_{\mathrm{C}} < V < V_{\mathrm{set}}$ will contribute to the exchange bias. The exchange bias in multigranular systems can be calculated as:
\begin{equation}
    H_{\mathrm{ex}} \propto \int^{V_{\mathrm{set}}(T)}_{V_{\mathrm{C}}(T)} f(V)dV,
    \label{eq:EBV}
\end{equation}
\noindent where the exchange bias is proportional to the number of grains between these critical volumes.

Although many models of exchange bias in polycrystalline thin films have been developed, all of these models assumed a distribution of set directions for the AFM grains\cite{Radu2008,Stiles1999ModelBilayers, VanDerHeijden1998InfluencesBilayers, Fulcomer1972ThermalCoupling,Malozemoff1987}. In this paper we continue with the natural model of Jenkins \etal~\cite{JenkinsEB2020} and instead of assuming an arbitrary distribution of set directions and anisotropies these will occur due to the naturally occurring distribution of the specific atomic configuration at the interface. We then perform large scale simulations of a multigranular $\gamma$-\IrMn / CoFe bilayer system investigating the setting of granular and continuous ferromagnetic layers, the computation of the exchange bias field and coercivity as a function of the system temperature and their dependence on the grain size dependence.

\section{Method}

Our simulations were performed using an atomistic spin model with the \textsc{vampire} software package~\cite{Evans2014}. The energetics of the system is described by the spin Hamiltonian:
\begin{eqnarray}
\mathscr{H} &=& -\sum_{i<j} \Jij \sS_i \cdot \sS_j - \frac{k_N}{2} \sum_{i \neq j}^z (\mathbf{S}_i \cdot  \mathbf{e}_{ij})^2 \nonumber \\
&& - \sum_i k_{\mathrm{u}} (\sS \cdot \mathbf{e}_z)^2 - \sum_i \smmu \sms_i \cdot \smB,
\label{eq:hamiltonian}
\end{eqnarray}
with $\sS_i$ describing the spin direction on site $i$, $k_N = -4.22 \times 10^{-22}$ is the \Neel pair anisotropy constant and $\mathbf{e}_{ij}$ is a unit vector from site $i$ to site $j$, $z$ is the number of nearest neighbours and \Jij is the exchange interaction. The effective exchange interactions (\Jij) were limited to nearest ($\Jnn = -6.4 \times 10^{-21}$ J/link) and next nearest ($\Jnnn = 5.1 \times 10^{-21}$ J/link) neighbours~\cite{Jenkins2018EnhancedFilms,Jenkins2019MagneticIrMn3}. 
The CoFe layer is simulated with a nearest neighbour approximation and a weak easy-plane anisotropy $k_{\mathrm{u}}$ to simulate the effects of the demagnetising field of a thin film. The exchange coupling across the FM/AFM interface is set at 1/5th of the bulk exchange values as calculated by \textit{ab-initio} methods~\cite{Szunyogh2011AtomisticInterface}. Spin Dynamics simulations were done solving the stochastic Landau-Lifshitz-Gilbert equation with a Heun numerical scheme~\cite{Ellis2015TheModels}. Our model naturally reproduces the low temperature ground state spin structures where the ordered alloy forms a triangular (T1) spin structure with an angle of 120 degrees between adjacent spins and the disordered alloy forms a tetrahedral (3Q) spin structure with 109.5 degrees between spins \cite{Jenkins2018EnhancedFilms} in agreement with previous neutron scattering experiments~\cite{Tomeno1999MagneticMn3Ir, Kohn2013TheExchange-bias.} and theoretical calculations~\cite{Szunyogh2011AtomisticInterface,Sakuma2003First-principlesAlloys,Hemmati2012MonteLattice}. The simulations also reproduce the \Neel ordering temperature of  730K for the disordered $\gamma$ phase~\cite{Yamaoka1974AntiferromagnetismAlloys}. The simulations were run in parallel on 400 cores to enable ns hysteresis loops, ensure a converged coercivity and value for the exchange bias (in the limit of critical damping).

\begin{figure}[!tb]
\centering
\includegraphics[width=8.5cm]{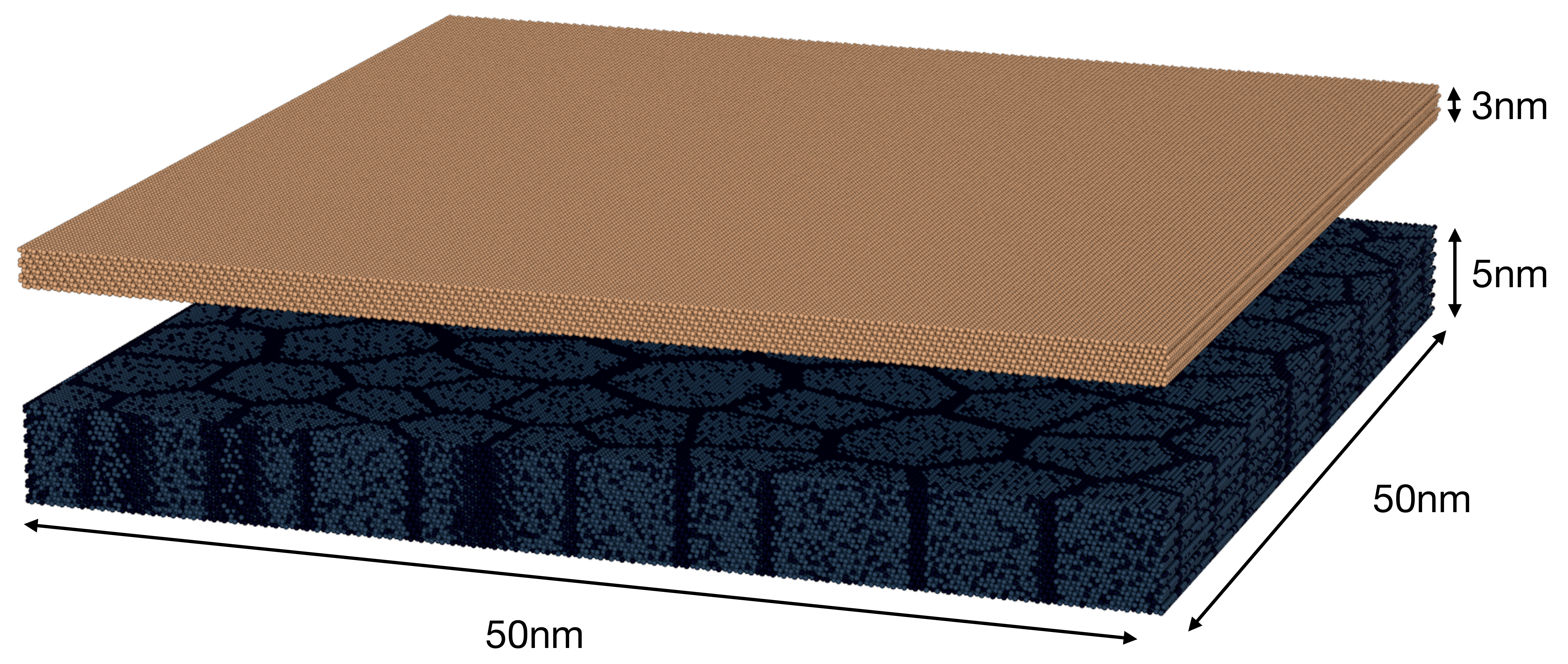}
\caption[Visualisation of the multigranular IrMn/CoFe bilayer structure]{\textbf{Visualisation of the multigranular IrMn/CoFe bilayer structure.} The CoFe is represented by gold spheres and lifted 5nm above the IrMn to show the multigranular structure below. The Ir is represented as black spheres and the Mn is dark blue. The system is 50nm by 50nm in size}
\label{fig:vis}
\end{figure}

\section{Results}
\begin{figure}[!tb]
\centering
\includegraphics[width=8.5cm]{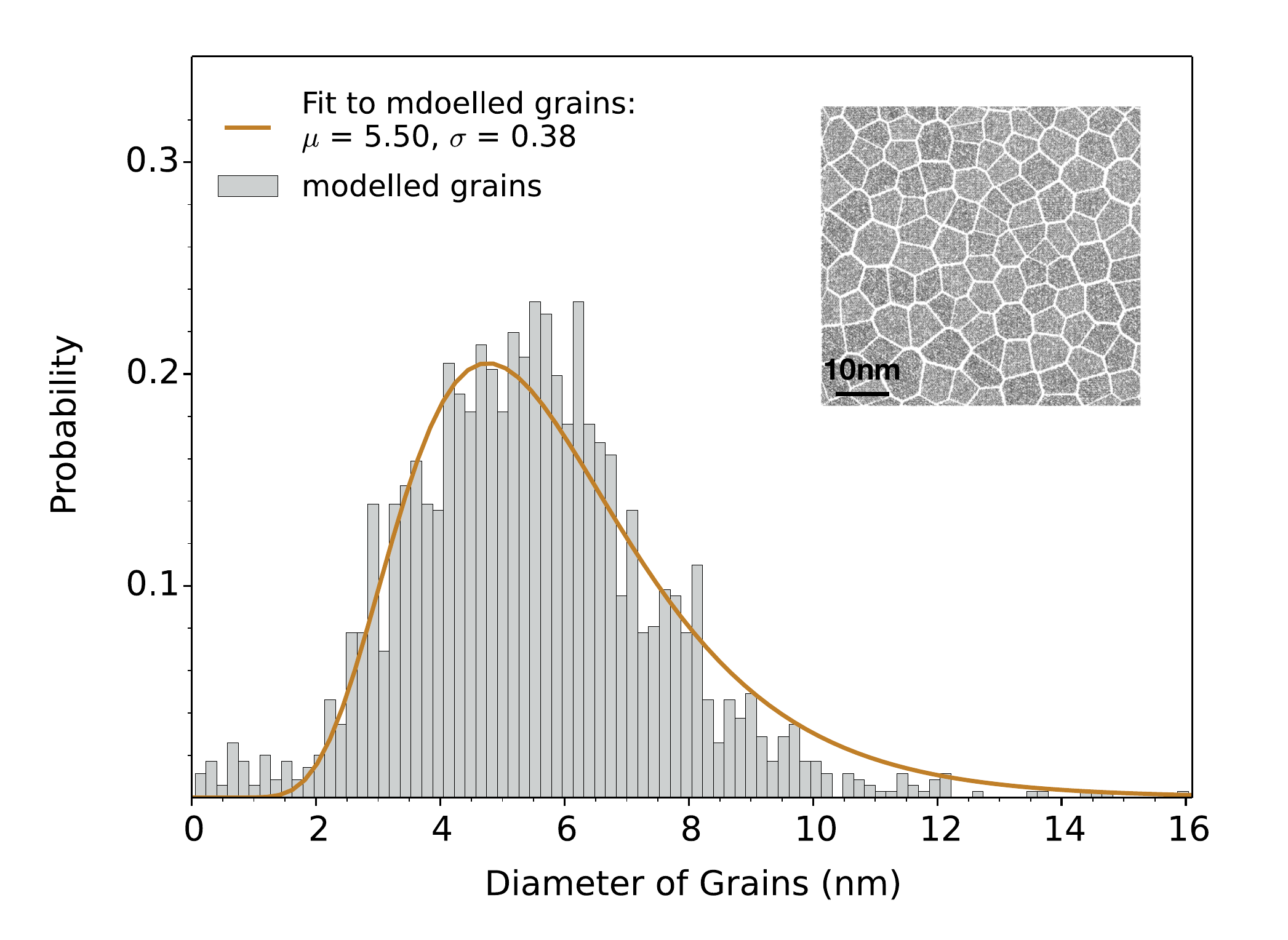}
\caption[The granular structure generated from the Poisson distribution]{\textbf{The granular structure generated from the Poisson distribution.}  The grain size distribution, the input median and standard deviation nearly match the output distribution. \textit{inset}. The granular structure generated. The grain shapes look realistic as do the distribution of grain sizes.  }
\label{fig:poisson_dist}
\end{figure}

To study the exchange bias effect, we couple a 5 nm thick \IrMn layer to a 3 nm thick ferromagnetic layer of CoFe to form a bilayer with a (111) out of the plane orientation of the \IrMn to reproduce the structure used in typical devices. The granular structure of the IrMn is created using the Poisson method~\cite{poisson}, where the seed points are generated using Poisson distribution and the grains are generated from this using a voronoi construction. There is no exchange across the grain boundaries, matching the deduction of related experimental measurements~\cite{OGrady2010AFilms}. The CoFe is modelled as a continuous film. The simulated structure was 50 nm $\times$ 50 nm and contains over 1.5 million atoms. A visualisation of this structure is shown in Fig.~\ref{fig:vis}. The initial grain distribution had a median grain size of 5.5 nm and a standard deviation of 0.37 and is shown in Fig.~\ref{fig:poisson_dist}. 

\begin{figure*}[!tb]
\centering
\includegraphics[width=0.8\textwidth]{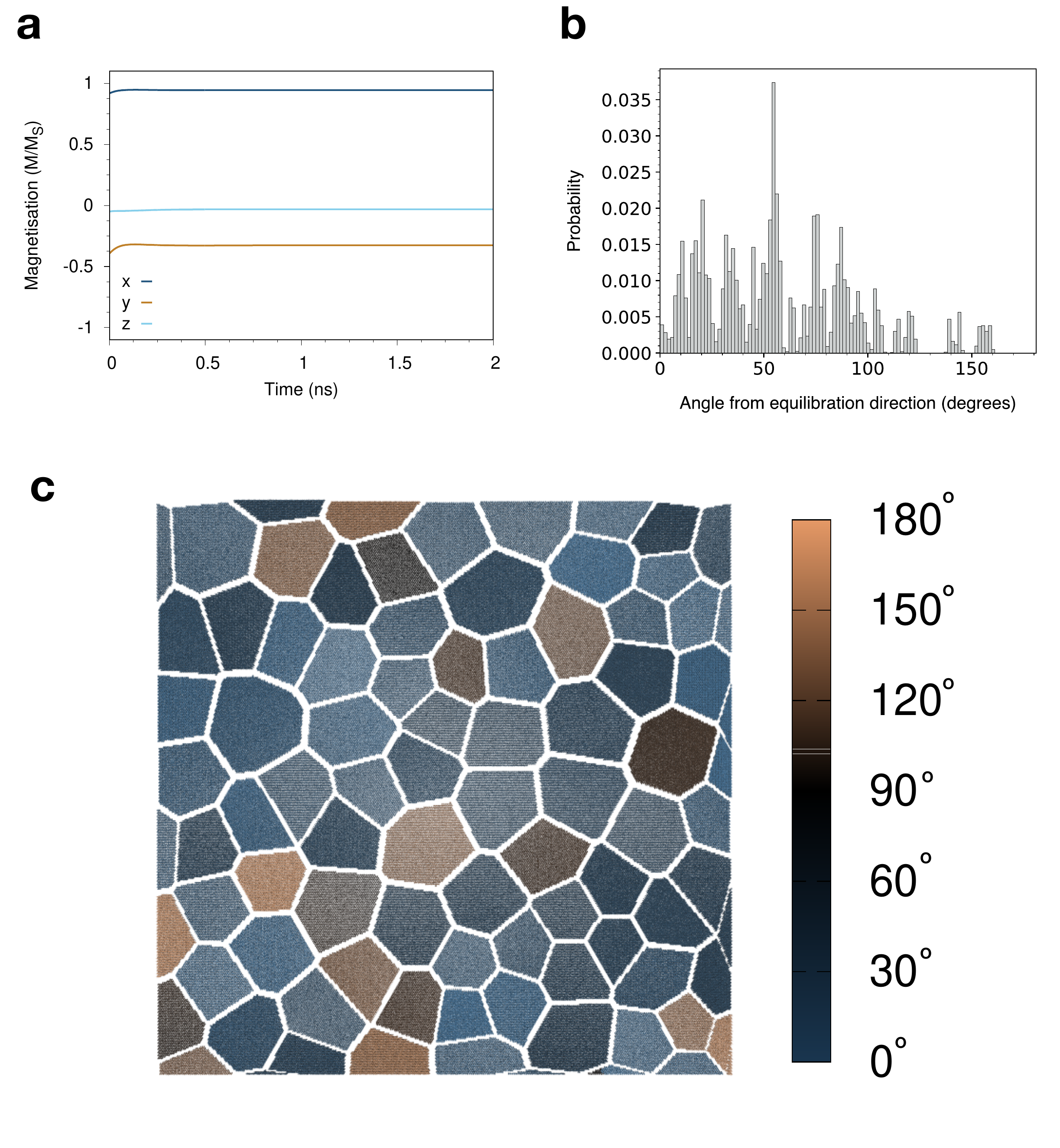}
\caption[The magnetisation direction throughout the equilibration stage of the simulation and direction of the net interface exchange field]{\textbf{The magnetisation direction throughout the equilibration stage of the simulation and direction of the net interface exchange field}. (a) The direction of the net magnetisation of the FM throughout the equilibration stage. The simulated magnetisation has remained along the (1,0,0) direction. (b) During the equilibration step the FM relaxes to its minimum energy position, as there is no applied field, the minimum energy occurs when the FM aligns with the interface moment of the AFM. In this simulation the FM also has a granular structure so each FM grain will follow the magnetisation of the interface field of the AFM below it. The angle of rotation away from the setting field direction is plotted on the histogram in (b) and shown schematically in (c). In (c) the colour in the diagram represents the angle to the setting field direction at the end of the equilibration simulation. Whilst most of the grains are have only canted 10 - 60$^\circ$ away from the setting field direction some of the grains are almost 150$^\circ$ away.}
\label{fig:eq_grains_uncoupled}
\end{figure*}

Experimentally, for exchange bias to occur the system needs to be field cooled under a high field~\cite{OGrady2010AFilms}. During this step the net direction of the uncompensated spins at the FM/AFM interface align with the field. Due to the small energy difference between the possible AFM ground states the switching takes place over a timescale of hours and if not done slowly enough the AFM will set along the wrong direction. As each grain is set individually, if not set correctly the exchange bias in the grains will set in random directions, giving no net exchange bias. Unfortunately, the time required to simulate hours is not computationally feasible. Instead the direction of the exchange bias is set by a simulated setting process as follows, where the field ($\mu_0 H_{\text{set}}$) was applied along the $x$-direction. 

A setting algorithm was developed to allow instantaneous setting of the system. The setting procedure forces the AFM to set with the direction of the net interface magnetisation along the direction of the setting field. From Jenkins \etal~\cite{JenkinsEB2020}, we know that the interface moment is caused by an imbalance of Mn atoms in each of the four sublattices, causing one sublattice to have a larger moment than the other four sublattices. In each grain the magnetisation of the AFM sublattices can lie along four possible directions. The setting procedure sets the magnetisation of the sublattice with the largest number of Mn atoms along the AFM magnetisation direction closest to the setting field direction. The other three sublattices are then set along the remaining three possible sublattice magnetisation directions. The setting of each sublattice along each direction was calculated from the geometry, and finally the magnetisation of the CoFe is set along the applied field direction.

\begin{figure*}[!tb]
\centering
\includegraphics[width=0.95\textwidth]{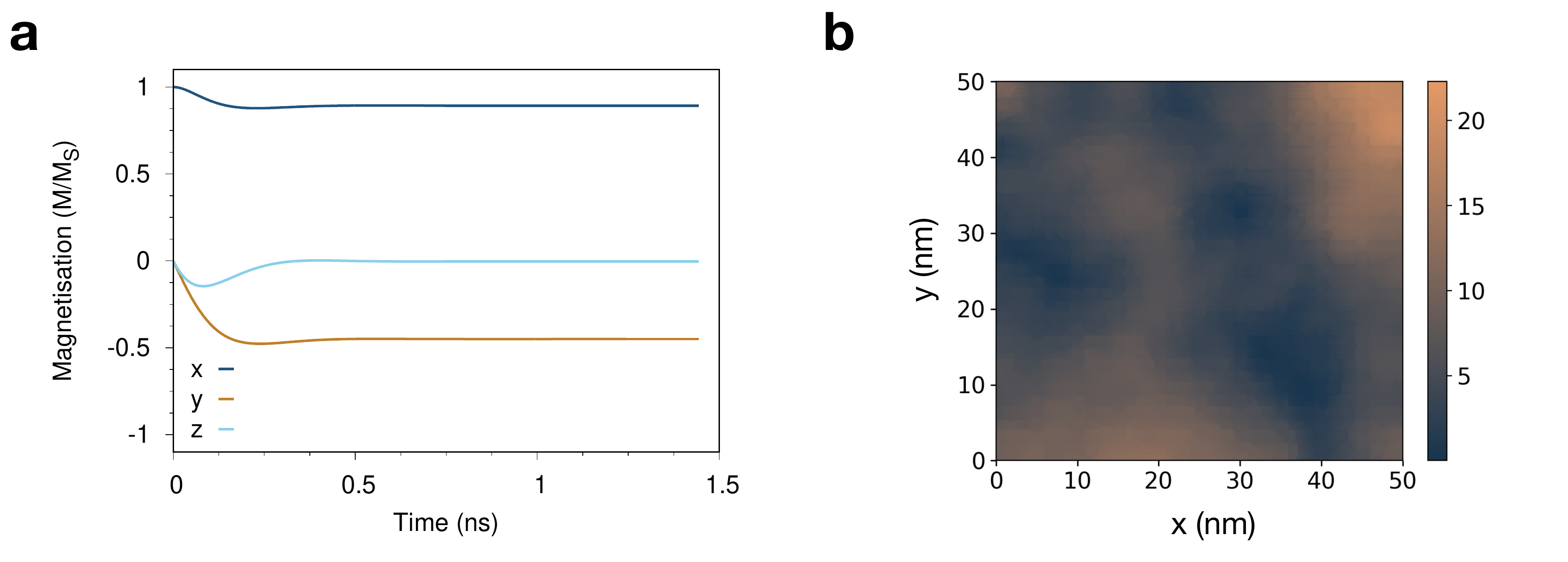}
\caption[The motion of the FM throughout the equilibration stage of the simulation]{\textbf{The motion of the FM throughout the equilibration stage of the simulation.} During the equilibration stage all external fields are removed and the only force the FM feels is from the AFM below. (a) The motion of the FM throughout the simulation. The FM cants slightly away from the setting field direction and into the direction of the interface moment from the AFM below. The direction of the interface field is only slightly away from the setting field direction. (b) The interface layer of the FM shows canting of up to 20 degrees and imprinting from the grains below.}
\label{fig:eq_grains_coupled}
\end{figure*}

\begin{figure}[!tb]
\centering
\includegraphics[width=8.5cm]{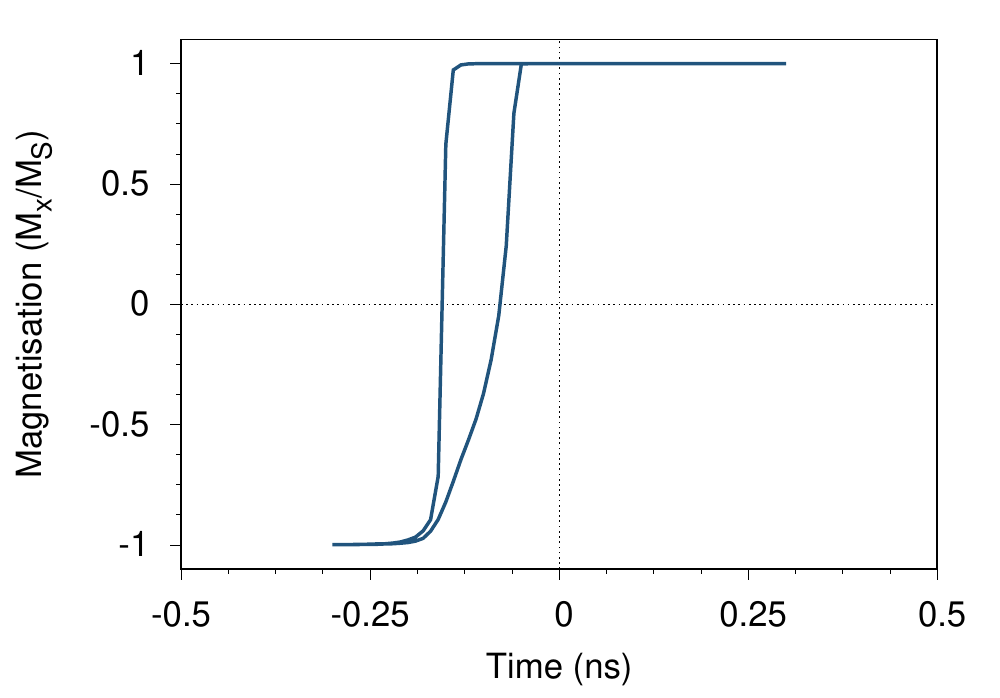}
\caption[Simulated hysteresis loop for a granular AFM]{\textbf{Simulated hysteresis loop for a granular AFM.} The hysteresis loop exhibits 0.12T of exchange bias.}
\label{fig:hy_grains}
\end{figure}

After setting, the field is removed and the system relaxes to an equilibrium state, where the FM cants slightly away from the setting field direction to (0.895,-0.440,0.001), approximately $26.5^\circ$ from the setting direction, as shown in Fig.~\ref{fig:eq_grains_uncoupled} (a). The canting is due to the distribution of set directions in the underlying AFM grains as shown in Fig.~\ref{fig:eq_grains_uncoupled} (b). The distribution has naturally occurred in the model due to the disorder present in the AFM structure. Most of the grains have been correctly set close to the setting field direction, however, a small proportion are incorrectly set and the magnetisation of the decoupled CoFe grains has canted up to 150$^\circ$. The incorrectly set grains are due to the complicated grain shapes, the fact that the strength and direction of the interface exchange field is a vector combination of the uncompensated interface spins, and in these more complicated structures the placement of the spins in the interface becomes more important and the setting procedure becomes less accurate.

It was proposed by Barker \etal~\cite{Barker2009ASensors} that at the FM/AFM interface the magnetic structure of the FM would show an imprint of the granular AFM magnetisation below. Here we simulate a continuous ferromagnetic CoFe layer where the individual grain level exchange bias is weight-averaged leading to a smaller deviation of the magnetization from the set direction, shown in Fig.~\ref{fig:eq_grains_coupled}(a). The spin structure in the interface later of the ferromagnet is shown in Fig.~\ref{fig:eq_grains_coupled}(b), where the colour of the spins represents the angle from the average FM direction. It shows the same imprinting pattern seen by Barker \etal.  Although individual grains cannot be seen the FM spins can be seen to rotate up to 20\% and the total FM magnetization $M/M_\mathrm{S}$ has reduced from 1 to 0.992. 

A hysteresis loop simulation was run along the equilibrated bias direction, between $\pm$ 0.3 T in steps of 0.01 T and at each step the system was time evolved for 200,000 1 fs time-steps. The hysteresis loop produced is shown in Fig. \ref{fig:hy_grains} and has an exchange bias of 0.12 T. Assuming a reduction in the exchange bias due to temperature effects, this value is close to typical experimental measurements~\cite{OGrady2010AFilms, Ohldag2003CorrelationSpins}. 

\begin{figure}[!tb]
\centering
\includegraphics[width=8.5cm]{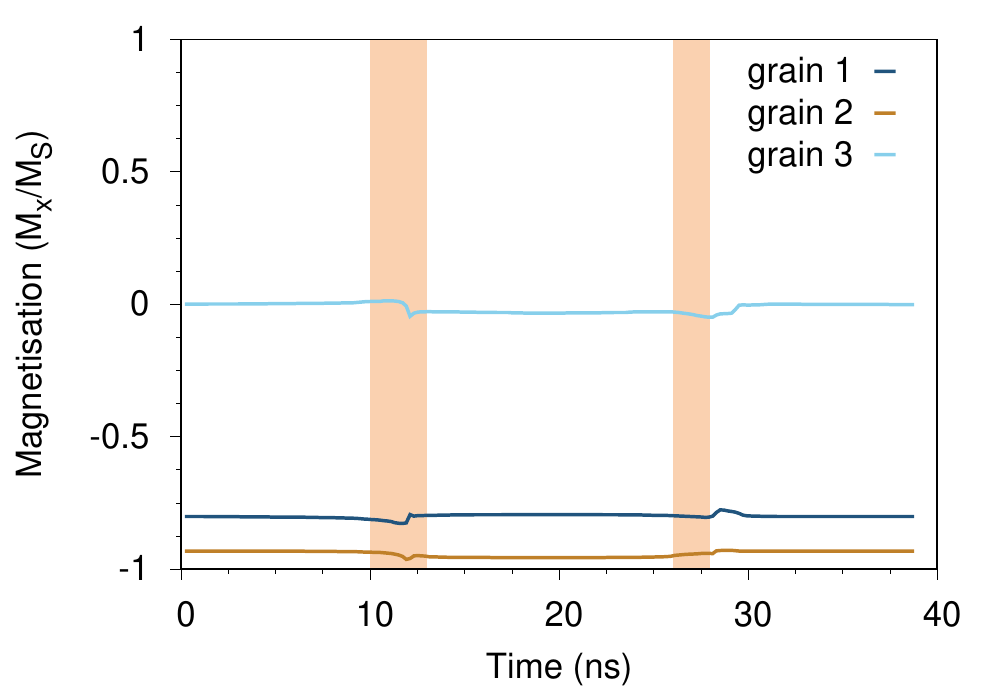}
\caption[Magnetisation along the x direction for sublattice 1 throughout the hysteresis loop for 3 different grains.]{\textbf{Magnetisation along the x direction for sublattice 1 throughout the hysteresis loop for 3 different grains.} Every grain of the AFM rotates at the same time, and the shaded rectangles show the points where the FM reverses magnetisation.}
\label{fig:AFM_rotate}
\end{figure}

The exchange bias is similar to the exchange bias found by Jenkins \etal~\cite{JenkinsEB2020} for a single grain system. In the multigrain system the exchange bias is an average of the individual grains explaining the similarity to the single grain result. The coercivity is 0.07 T, which is much smaller than the single grain coercivity from Jenkins \etal~\cite{JenkinsEB2020} of 0.13 T. There are two possible reasons for this decrease in coercivity. Firstly, there is now an angular dependence to the magnetisation of the grains and an increase in the angle between the field and the easy axis reduces the coercivity, much in the same way as for a ferromagnetic system~\cite{Stoner1947InterpretationMaterials}. Secondly, the larger ferromagnetic system now rotates with non-coherent rotation which also reduces the coercivity.

The exchange bias of the system is defined from Eq.~\ref{eq:EBV} as the integral of all the grains between $V_C$ and $V_{\mathrm{set}}$. However, as this hysteresis loop was simulated at $T=0$ K, even the smallest grains will be stable and as we have forced the grains to correctly set, the exchange bias should be the integral over all grains. As the exchange coupling of the FM layer is much stronger than the interface exchange coupling the FM will only rotate when the field is higher than the net field from the AFM. It can be observed that every AFM grain flips at the same time slightly after the FM as shown in Fig.~\ref{fig:AFM_rotate}. The FM-AFM reversible moment in all grains can therefore be said to rotate coherently with the FM due to the large exchange field. We note that the reversible component does not contribute to the exchange bias but does contribute to the coercivity. The natural misalignment of the individual exchange bias directions in the grains and coherent switching therefore explain the reduction in the coercivity compared to that of a single grain.

\subsection{Temperature dependence of exchange bias}
\begin{figure*}[!tb]
\centering
\includegraphics[width=0.95\textwidth]{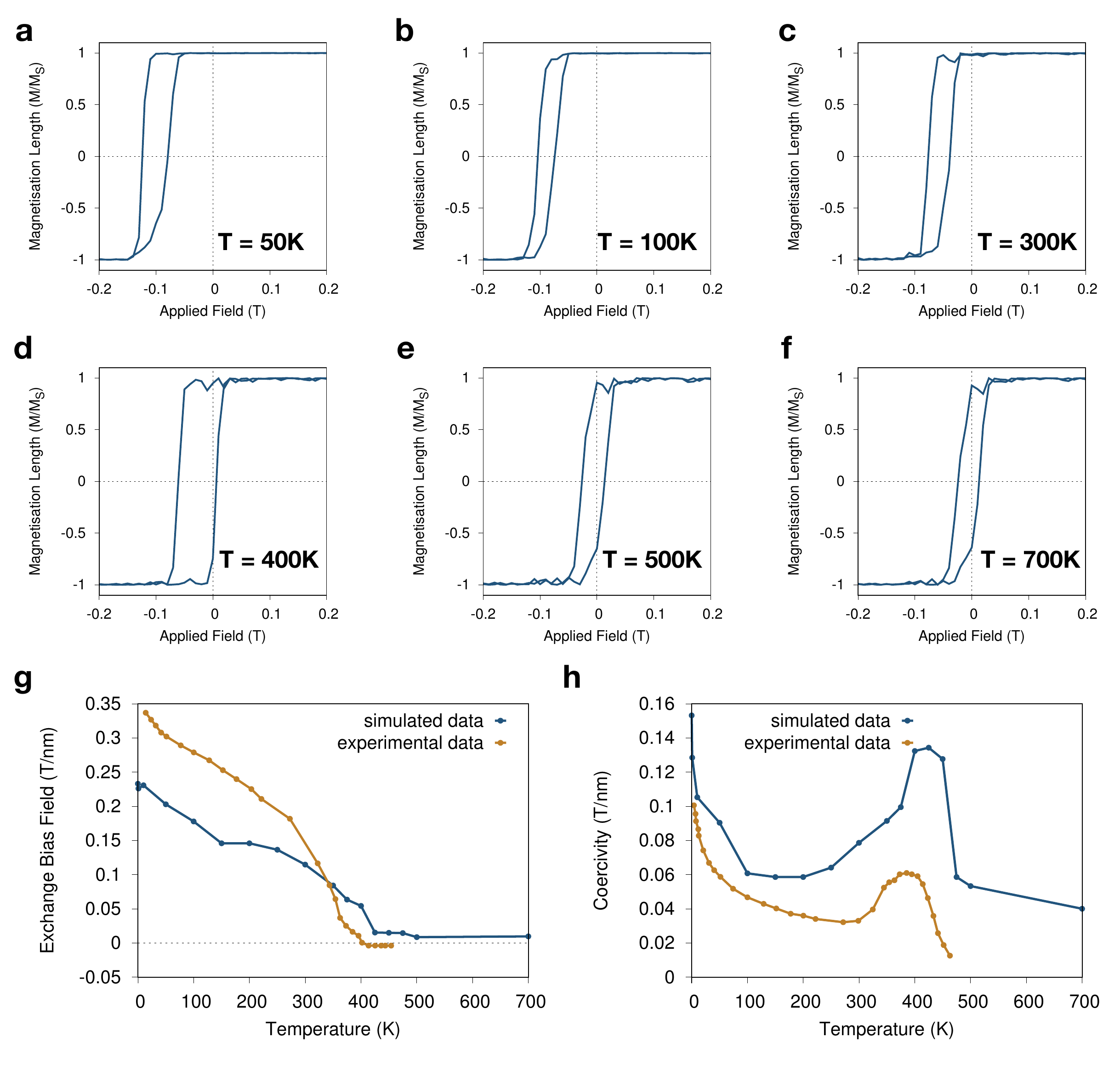}
\caption[Temperature dependence of exchange bias]{\textbf{Temperature dependence of exchange bias} Simulated hysteresis loop for a granular AFM at (a) 50K , (b) 100K, (c) 300K, (d) 400K, (e) 500K and (f) 700K. (g) The simulated temperature dependence of the exchange bias and (h) the coercivity, compared with the experimental results of Ali \etal~\cite{Ali2003AntiferromagneticSystem}. The simulated exchange bias decreases with temperature as does the experimental result.}
\label{fig:temp}
\end{figure*}

To investigate the temperature dependence of the exchange bias, the temperature of the hysteresis loop simulation was systematically varied from 0K to 700K. The simulated hysteresis loops for 50K, 100K, 300K, 400K, 500K and 700K in Fig. \ref{fig:temp} and the computed exchange bias and coercivity are plotted in Fig.~\ref{fig:temp} (g) and (h) respectively. At 300K the exchange bias is 0.06T, $\sim 50$\% of the 0K value. Our multigranular system contains small ($\sim 2$ nm) grains as shown in Fig.~\ref{fig:poisson_dist}. At 300K these small grains will be completely thermally unstable as the temperature is larger than the blocking temperature, causing the decrease in the exchange bias. 

\begin{figure*}[!tb]
\centering
\includegraphics[width=0.99\textwidth]{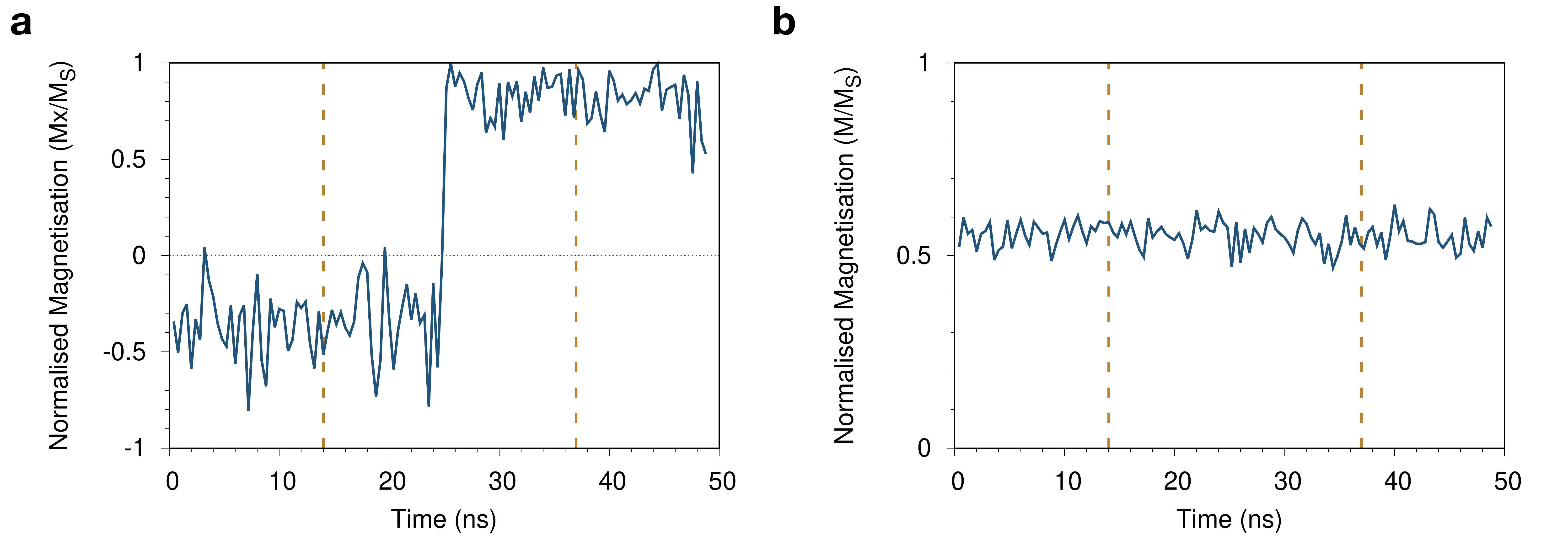}
\caption[Magnetisation along x of one AFM sublattice in one grain, it can be seen to rotate at negative saturation of the FM. ]{\textbf{Magnetisation along x of one AFM sublattice in one grain, it can be seen to rotate at negative saturation of the FM.} (a) The magnetisation rotates between the positive and negative exchange bias directions. The points the FM flips are outlined by the gold dashed lines. (b) The magnetisation length remains constant suggesting the grain flips coherently and does not form domains.}
\label{fig:EBHC_temp}
\end{figure*}

The exchange bias decreases with temperature and the coercivity initially decreases but then
increases to a peak at about 400K - 450K. The peak in the coercivity matches the temperature that the exchange bias decreases to zero, which corresponds to the blocking temperature of the AFM. The temperature dependence of the exchange bias and the coercivity was experimentally measured in IrMn/CoFe systems for varying thicknesses of CoFe by Ali \etal~\cite{Ali2003AntiferromagneticSystem}. The experimental data shows exactly the same trend as the simulated results with the exchange bias decreasing and the coercivity having a peak at 400K - 450K. At 400K the exchange bias disappears because the system has reached the blocking temperature and the grains now have enough thermal energy to rotate between ground states. But why does this cause a large increase in the coercivity if there is no exchange bias? To investigate this, the change in magnetisation of the AFM in each grain was observed throughout the hysteresis loop at the blocking temperature (400K). The magnetisation along $x$ of one of the AFM sublattices in one grain is shown in Fig.~\ref{fig:EBHC_temp}(a), and the magnetisation of the AFM can be seen to reverse after the FM reverses. The magnetisation then remains along this new direction. The magnetisation length is shown in Fig.  \ref{fig:EBHC_temp}(b), showing that the magnetisation length remains constant at approximately 0.6 - which is the value of $M/M_s$ at 400K for bulk IrMn$_3$. This suggests that the IrMn$_3$ is rotating coherently and not breaking up into domains. This behaviour is observed in a large proportion of the grains. 

The flipping of the AFM means that instead of the AFM adding a unidirectional anisotropy now it adds a uniaxial anisotropy. This means it adds exchange bias in both directions, as after flipping, the exchange bias is now in the opposite direction and has been thermally reset during the hysteresis loop. This thermal resetting causes the increase in coercivity even with no exchange bias. 

The experimental coercivity has a slightly smaller magnitude than the simulated data. Here we have taken a measurement from only the first hysteresis loop, however, it is well known from Sharrock's law that the coercivity is time-dependent~\cite{Sharrock1994TimeMedia} and the experimental results are done over seconds whereas ours are done over ns so more grains will flip earlier in the experimental measurements than in our simulations. 

\subsection{The grain size dependence of exchange bias}

\begin{figure*}[!tb]
\centering
\includegraphics[width=0.99\textwidth]{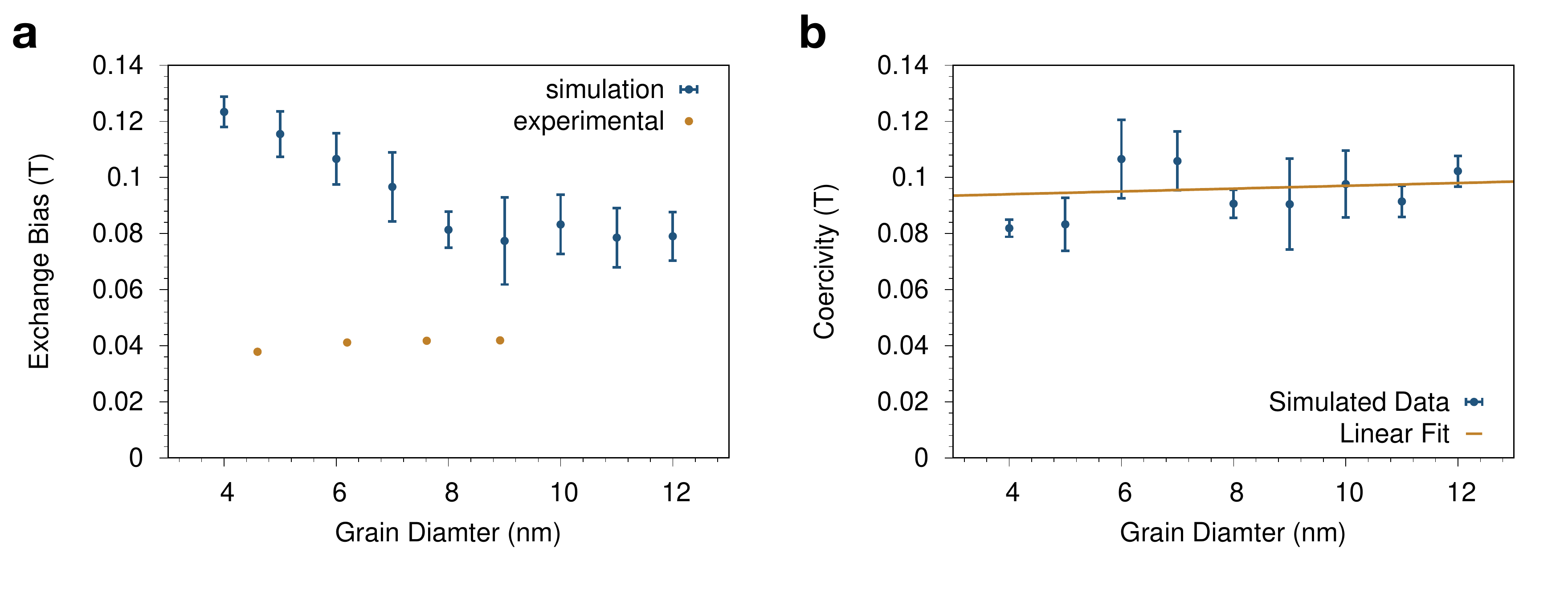}
\caption[The simulated grain size dependence of the exchange bias and coercivity at $T=0$K compared to experimental results]{\textbf{The simulated grain size dependence of the exchange bias and coercivity at $T=0$K compared to experimental results~\cite{OGrady2010AFilms}.} (a) The exchange bias has a maximum value in 4 nm grains in contradiction with the experiments. The experimental results were measured at $T=300$K, explaining the difference in magnitude between the two datasets. The fit to the experimental data is taken from~\cite{OGrady2010AFilms}(b). The coercivity of the hysteresis loop seems to be unrelated to the grain diameter, as shown by the linear fit with a gradient of only 0.0005T.}
\label{fig:EBHC_grain0}
\end{figure*}

Real devices will have a distribution of grain sizes depending on the growth techniques. To investigate the role of the granular distribution the system dimensions the median grain size was varied from 4nm - 12nm. The standard deviation of the grain size distribution was kept constant at 0.37. Five simulations were run for each grain diameter each with different random numbers used to generate the granular structure so an average exchange bias could be calculated.

\begin{figure}[!tb]
\centering
\includegraphics[width=8.5cm]{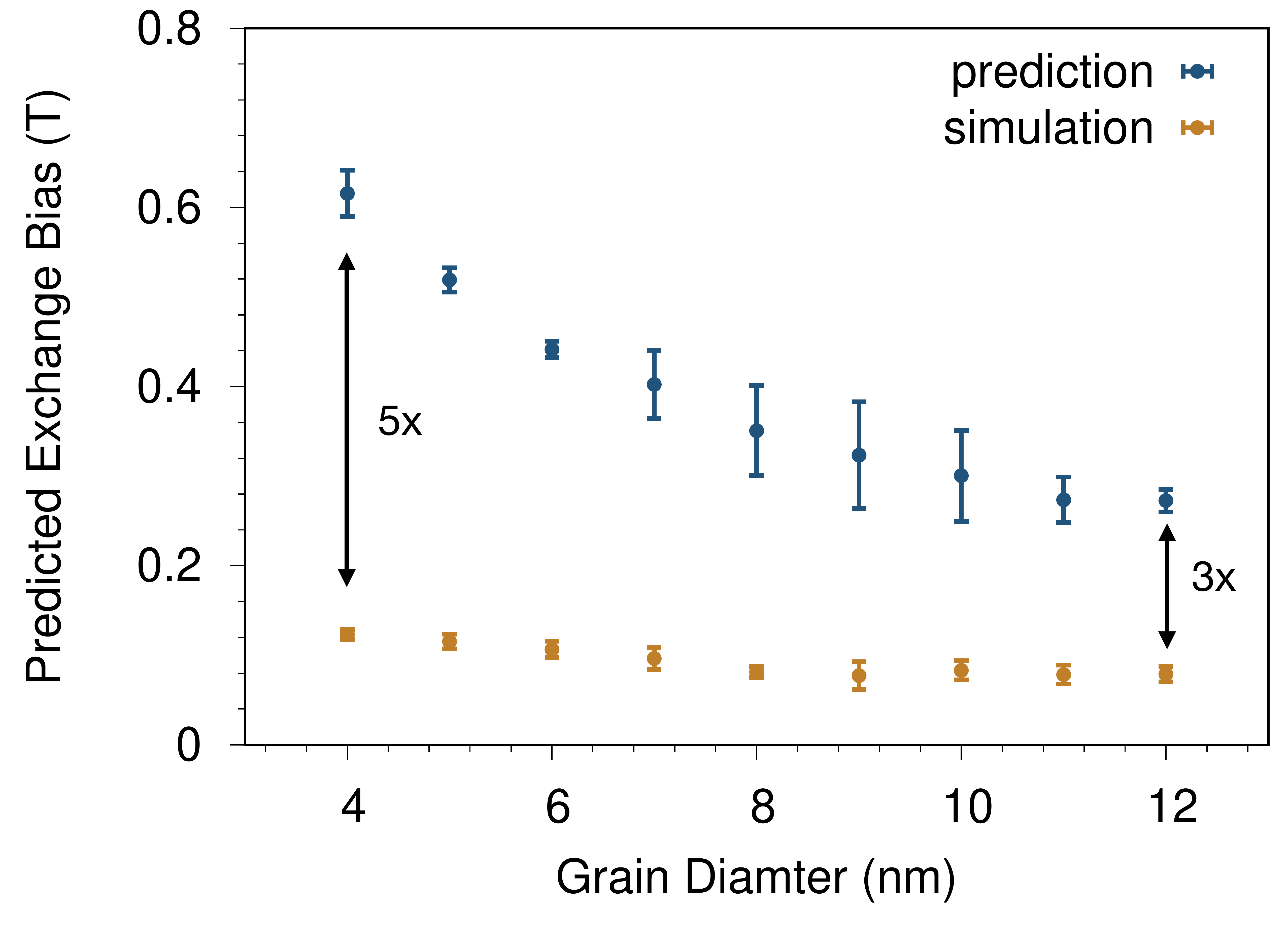}
\caption[Predicted exchange bias in the multi-granular system for different grain sizes]{\textbf{Predicted exchange bias in the multi-granular system for different grain sizes}. The exchange bias decreases with the grain diameter as observed in our simulations. }
\label{fig:Nun}
\end{figure}

The hysteresis loop simulations were initially run at 0K and the average exchange bias was calculated from the five simulations, as plotted in Fig. \ref{fig:EBHC_grain0} with the experimental results from O'Grady \textit{et al}. The exchange bias has a maximum for smaller grain sizes, because the smaller the grain size the larger the statistical imbalance between the number of spins in each sublattice~\cite{JenkinsEB2020}. In reality, with temperature the small grains would become superparamgnetic and not contribute to the exchange bias as in equation \ref{eq:EBV}. The number of uncompensated spins in each grain ($n_{un}$) can be predicted as in Jenkins \etal~\cite{JenkinsEB2020}. The number of uncompensated spins can be calculated for each grain, then summed to calculate the number of uncompensated spins for the entire system. From the number of uncompensated spins the exchange bias can be calculated from equation \ref{eq:EBnun}. 

The predicted exchange bias for each grain size averaged over the five systems is plotted in Fig. \ref{fig:Nun}. It shows the same pattern as shown in Fig. \ref{fig:EBHC_grain0}(a) but the predictions are about five times higher than the simulated values for small grain diameters and about three times higher for large grain diameters. The reduction from the predictions is likely due to the unset grains as shown in Fig. \ref{fig:eq_grains_uncoupled}, and from the presence of spins at the edges of the grains. This effect is larger for smaller grain sizes, due to the increased edge to volume ratio, explaining the larger difference from the prediction for smaller grains than larger grains. 

\begin{figure}[!tb]
\centering
\includegraphics[width=8cm]{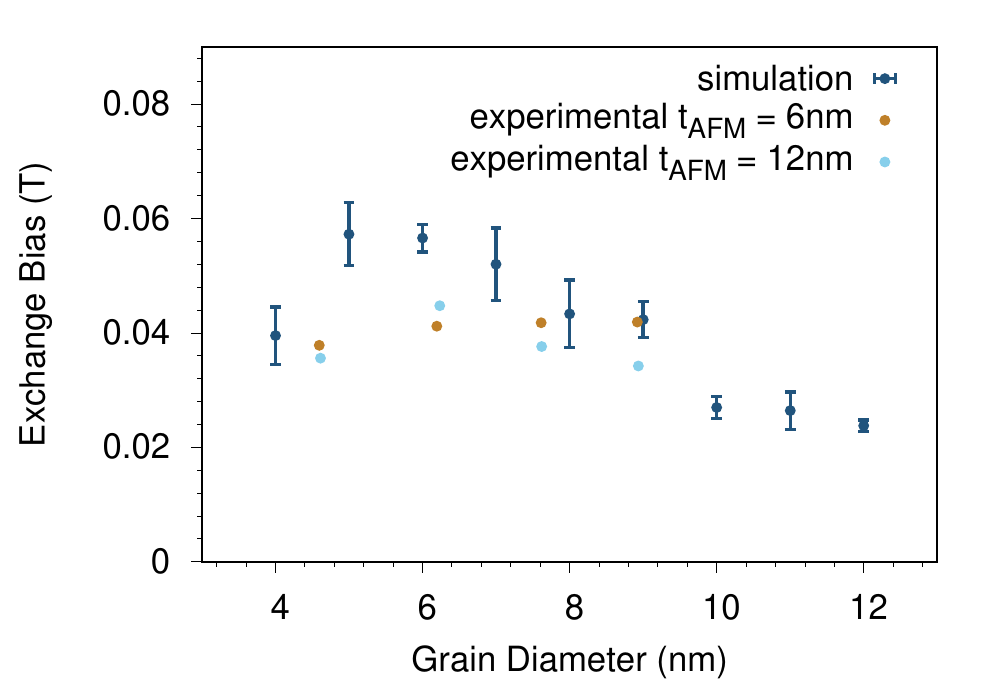}
\caption[The simulated grain size dependence of the exchange bias and coercivity at $T=300$K compared to experimental results]{\textbf{The simulated grain size dependence of the exchange bias and coercivity at $T=300$K compared to experimental results~\cite{OGrady2010AFilms}}. The dependence of the exchange bias with grain size at 300K. The experimental results for a AFM thickness of 6nm and 12nm are shown, our simulation behaves more like a 12nm system than a 6nm system even though our AFM thickness was only 5nm.}
\label{fig:EB_grain300}
\end{figure}

The exchange bias is also about five times higher than the experimental results, because our simulations were run at 0K. We expect that at room temperature the exchange bias of the small grain sizes will decrease because the smaller grains will become thermally unstable. The coercivity at $T=0$ K is plotted in Fig. \ref{fig:EBHC_grain0}(b) and is unrelated to the grain size. Here the athermal coercivity is entirely due to the reversible moment in the antiferromagnet interface which is naturally independent of the grain size.

To consider the effects of grain size at higher temperatures, we now compute the same data at $T = 300$ K. With the inclusion of thermal effects the exchange bias for low diameter grains has decreased, as shown in Fig. \ref{fig:EB_grain300}. The smallest grains are unstable and no longer contribute to the exchange bias. The results are plotted against experimental data for film thicknesses of 6 nm and 12 nm~\cite{OGrady2010AFilms}, where the thickness of the FM shifts the peak in the exchange bias as the peak is proportional to $KV/k_BT$~\cite{OGrady2010AFilms}. The simulated data has a maximum at a 6nm diameter as does the experimental data for a 12nm thick AFM. At 300K the exchange bias has dramatically reduced for large grain sizes to 25\% of the 0K value whereas for a 6nm grain diameter the reduction is only about 50\%. The 300K trend matches the trend seen experimentally but it was predicted to be due to the fact that the large grains are not set correctly during the setting process. This cannot explain the reduction in exchange bias seen here from 0K to 300K as the grains were set exactly the same in both simulations. One reason might be that there are too few grains in the simulations as for 12 nm grains a 50 nm $\times$ 50 nm system will only fit in about 20 grains meaning any unset grains will drastically reduce the exchange bias. A more quantitative comparison may be possible in future with increased availability of computing power. 


\section{Conclusion}
In conclusion, we have deveoped a large scale atomistic model of a multigranular \IrMn / CoFe bilayer, naturally including the non-collinear nature of the IrMn layer. The 50 nm size of the simulated system is comparable to realistic devices and contains over 1.5 million atoms. The model includes a natural distribution for the exchange bias directions of the AFM grains due to the statistical imbalance in the number of AFM spins in each sublattice of the AFM as predicted by Jenkins \etal~\cite{JenkinsEB2020}. Our simulations give realistic values of exchange bias and both the temperature and grain size dependence qualitatively match previous experimental measurements. Our results demonstrate two distinct contributions to the coercivity in exchange bias systems arising from the athermal reversible component of the interfacial spins, and a thermal contribution due to the thermally assisted flipping of grains close to the blocking temperature. At 300K, we found a peak in the exchange bias for a median grain size of 6nm, while for smaller grains the exchange bias has decreased due to the increased thermal instabilities. Future work will consider the effects of interlayer intermixing and the origin of the athermal training effect. The model provides a more comprehensive understanding of the origins of coercivity and exchange bias in multigranular systems, and particularly in their thermal stability for different grain sizes. The increased understanding will provide possibilities for optimisation of exchange biased systems and the possible development of neuromorphic~\cite{GrollierNelec2020} and antiferromagnetic spintronic \cite{BaltzRMP2018,JungwirthNnano2016} devices. Our model also forms the basis of nanoscale antiferromagnetic spintonic device modelling including dynamics \cite{JenkinsJAP2020} that may provide further insights on electrically induced antiferromagnetic switching~\cite{LinNatMat2019,KimAPL2019} and the operation of neuromorphic computing devices~\cite{FukamiNatMat2016}.

\section*{acknowledgements}
The authors are grateful for provision of computer time on the \textsc{viking} cluster at the University of York. SJ is grateful for the provision of a PhD studentship from Seagate Technology (Derry). This work used the ARCHER UK National Supercomputing Service (http://www.archer.ac.uk) using code enhancements implemented and funded under the ARCHER embedded CSE programme (eCSE0709).

\section*{author contributions}
SJ  developed  the  algorithm,  generated  the  data,  analysed the results and drafted the manuscript. All authors contributed to the discussion of results and the final manuscript.

\bibliography{references,library}

\end{document}